\documentstyle[twoside,psfig]{article}

\catcode`\@=11
\long\def\@makefntext#1{
\protect\noindent \hbox to 3.2pt {\hskip-.9pt  
$^{{\eightrm\@thefnmark}}$\hfil}#1\hfill}		

\def\@makefnmark{\hbox to 0pt{$^{\@thefnmark}$\hss}}	
	
\def\ps@myheadings{\let\@mkboth\@gobbletwo
\def\@oddhead{\hbox{}
\rightmark\hfil\eightrm\thepage}   
\def\@oddfoot{}\def\@evenhead{\eightrm\thepage\hfil
\leftmark\hbox{}}\def\@evenfoot{}
\def\sectionmark##1{}\def\subsectionmark##1{}}

\oddsidemargin=\evensidemargin
\newcounter{sectionc}\newcounter{subsectionc}\newcounter{subsubsectionc}
\renewcommand{\section}[1] {\vspace{12pt}\addtocounter{sectionc}{1} 
\setcounter{subsectionc}{0}\setcounter{subsubsectionc}{0}\noindent 
	{\tenbf\thesectionc. #1}\par\vspace{5pt}}
\renewcommand{\subsection}[1] {\vspace{12pt}\addtocounter{subsectionc}{1} 
	\setcounter{subsubsectionc}{0}\noindent 
	{\bf\thesectionc.\thesubsectionc. {\kern1pt \bfit #1}}\par\vspace{5pt}}
\renewcommand{\subsubsection}[1] {\vspace{12pt}\addtocounter{subsubsectionc}{1}
	\noindent{\tenrm\thesectionc.\thesubsectionc.\thesubsubsectionc.
	{\kern1pt \tenit #1}}\par\vspace{5pt}}
\newcommand{\nonumsection}[1] {\vspace{12pt}\noindent{\tenbf #1}
	\par\vspace{5pt}}

\newcounter{appendixc}
\newcounter{subappendixc}[appendixc]
\newcounter{subsubappendixc}[subappendixc]
\renewcommand{\thesubappendixc}{\Alph{appendixc}.\arabic{subappendixc}}
\renewcommand{\thesubsubappendixc}
	{\Alph{appendixc}.\arabic{subappendixc}.\arabic{subsubappendixc}}

\renewcommand{\appendix}[1] {\vspace{12pt}
        \refstepcounter{appendixc}
        \setcounter{figure}{0}
        \setcounter{table}{0}
        \setcounter{lemma}{0}
        \setcounter{theorem}{0}
        \setcounter{corollary}{0}
        \setcounter{definition}{0}
        \setcounter{equation}{0}
        \renewcommand{\thefigure}{\Alph{appendixc}.\arabic{figure}}
        \renewcommand{\thetable}{\Alph{appendixc}.\arabic{table}}
        \renewcommand{\theappendixc}{\Alph{appendixc}}
        \renewcommand{\thelemma}{\Alph{appendixc}.\arabic{lemma}}
        \renewcommand{\thetheorem}{\Alph{appendixc}.\arabic{theorem}}
        \renewcommand{\thedefinition}{\Alph{appendixc}.\arabic{definition}}
        \renewcommand{\thecorollary}{\Alph{appendixc}.\arabic{corollary}}
        \renewcommand{\theequation}{\Alph{appendixc}.\arabic{equation}}
        \noindent{\tenbf Appendix \theappendixc #1}\par\vspace{5pt}}
\newcommand{\subappendix}[1] {\vspace{12pt}
        \refstepcounter{subappendixc}
        \noindent{\bf Appendix \thesubappendixc. {\kern1pt \bfit #1}}
	\par\vspace{5pt}}
\newcommand{\subsubappendix}[1] {\vspace{12pt}
        \refstepcounter{subsubappendixc}
        \noindent{\rm Appendix \thesubsubappendixc. {\kern1pt \tenit #1}}
	\par\vspace{5pt}}

\newcommand{\textlineskip}{\baselineskip=13pt}
\newcommand{\smalllineskip}{\baselineskip=10pt}

\def\eightcirc{
\begin{picture}(0,0)
\put(4.4,1.8){\circle{6.5}}
\end{picture}}
\def\eightcopyright{\eightcirc\kern2.7pt\hbox{\eightrm c}} 

\newcommand{\copyrightheading}[1]
	{\vspace*{-2.5cm}\smalllineskip{\flushleft
	{\footnotesize Modern Physics Letters A #1}\\
	{\footnotesize $\eightcopyright$\, World Scientific Publishing
	 Company}\\
	 }}

\def\abstracts#1#2#3{{
	\centering{\begin{minipage}{4.5in}\footnotesize\baselineskip=10pt
	\parindent=0pt #1\par 
	\parindent=15pt #2\par
	\parindent=15pt #3
	\end{minipage}}\par}} 

\def\keywords#1{{
	\centering{\begin{minipage}{4.5in}\footnotesize\baselineskip=10pt
	{\footnotesize\it Keywords}\/: #1
	 \end{minipage}}\par}}

\newcommand{\bibit}{\nineit}

\renewenvironment{thebibliography}[1]
	{\frenchspacing
	 \ninerm\baselineskip=11pt
	 \begin{list}{\arabic{enumi}.}
        {\usecounter{enumi}\setlength{\parsep}{0pt}     
	 \setlength{\leftmargin 12.7pt}{\rightmargin 0pt} 
         \setlength{\itemsep}{0pt} \settowidth
	{\labelwidth}{#1.}\sloppy}}{\end{list}}

\newcounter{itemlistc}
\newcounter{romanlistc}
\newcounter{alphlistc}
\newcounter{arabiclistc}

\newcommand{\fcaption}[1]{
        \refstepcounter{figure}
        \setbox\@tempboxa = \hbox{\footnotesize Fig.~\thefigure. #1}
        \ifdim \wd\@tempboxa > 5in
           {\begin{center}
        \parbox{5in}{\footnotesize\smalllineskip Fig.~\thefigure. #1}
            \end{center}}
        \else
             {\begin{center}
             {\footnotesize Fig.~\thefigure. #1}
              \end{center}}
        \fi}

\newcommand{\tcaption}[1]{
        \refstepcounter{table}
        \setbox\@tempboxa = \hbox{\footnotesize Table~\thetable. #1}
        \ifdim \wd\@tempboxa > 5in
           {\begin{center}
        \parbox{5in}{\footnotesize\smalllineskip Table~\thetable. #1}
            \end{center}}
        \else
             {\begin{center}
             {\footnotesize Table~\thetable. #1}
              \end{center}}
        \fi}

\def\@citex[#1]#2{\if@filesw\immediate\write\@auxout
	{\string\citation{#2}}\fi
\def\@citea{}\@cite{\@for\@citeb:=#2\do
	{\@citea\def\@citea{,}\@ifundefined
	{b@\@citeb}{{\bf ?}\@warning
	{Citation `\@citeb' on page \thepage \space undefined}}
	{\csname b@\@citeb\endcsname}}}{#1}}

\newif\if@cghi
\def\cite{\@cghitrue\@ifnextchar [{\@tempswatrue
	\@citex}{\@tempswafalse\@citex[]}}
\def\citelow{\@cghifalse\@ifnextchar [{\@tempswatrue
	\@citex}{\@tempswafalse\@citex[]}}
\def\@cite#1#2{{$\null^{#1}$\if@tempswa\typeout
	{IJCGA warning: optional citation argument 
	ignored: `#2'} \fi}}

\def\pmb#1{\setbox0=\hbox{#1}
	\kern-.025em\copy0\kern-\wd0
	\kern.05em\copy0\kern-\wd0
	\kern-.025em\raise.0433em\box0}

\def\fnt#1#2{\footnotetext{\kern-.3em
	{$^{\mbox{\scriptsize #1}}$}{#2}}}

\def\ps@myheadings{%
    \let\@oddfoot\@empty\let\@evenfoot\@empty
    \def\@evenhead{\slshape\leftmark\hfil}
    \def\@oddhead{\hfil{\slshape\rightmark}}
    \let\@mkboth\@gobbletwo
    \let\sectionmark\@gobble
    \let\subsectionmark\@gobble
    }
\font\tenrm=cmr10
\font\tenit=cmti10 
\font\tenbf=cmbx10
\font\bfit=cmbxti10 at 10pt
\font\ninerm=cmr9
\font\nineit=cmti9

\font\eightrm=cmr8

\def\qed{\hbox{${\vcenter{\vbox{			
   \hrule height 0.4pt\hbox{\vrule width 0.4pt height 6pt
   \kern5pt\vrule width 0.4pt}\hrule height 0.4pt}}}$}}

\def\be {\begin{equation}}
\def\ee {\end{equation}}
\def\ba {\begin{eqnarray}}
\def\ea {\end{eqnarray}}

\def\m  {\mu}

\def\o  {\omega}

\def\p  {\pi}
\def\P  {\Pi}

\def\la {\label}
\def\le {\left}
\def\ri {\right}
\def\pa {\partial}
\def\f {\frac}

\def\bi {\begin{itemize}}
\def\ei {\end{itemize}}

\def\da {\dagger}
\def\ha {\hat}
\def\car {{\cal R}}
\begin{document}
\setlength{\textheight}{7.7truein}  

\thispagestyle{empty}

\markboth{\protect{\footnotesize\it Black hole area quantization $\ldots$}}{\protect{\footnotesize\it $\ldots$  Das, Ramadevi and Yajnik}}

\normalsize\textlineskip

\setcounter{page}{1}

\copyrightheading{}			

\vspace*{0.88truein}

\centerline
{\bf BLACK HOLE AREA QUANTIZATION\footnote{Presented 
by U. A. Yajnik at IGQR workshop, IUCAA, December 2001}}

\vspace*{0.4truein}
\centerline{\footnotesize SAURYA DAS\footnote{saurya@theory.uwinnipeg.ca
}}
\baselineskip=12pt
\centerline{\footnotesize\it Physics Department, The University of 
Winnipeg and}
\baselineskip=10pt
\centerline{\footnotesize\it Winnipeg Institute for Theoretical Physics,}
\centerline{\footnotesize\it 515 Portage Avenue, Winnipeg, Manitoba R3B 2E9,  
Canada}
\vspace*{12pt}

\centerline{\footnotesize P. RAMADEVI and U. A. YAJNIK
\footnote{ramadevi, yajnik@phy.iitb.ac.in}}
\baselineskip=12pt
\centerline{\footnotesize\it Physics Department, Indian Institute 
of Technology, Bombay}
\baselineskip=10pt
\centerline{\footnotesize\it Mumbai 400\thinspace076, India}
\vspace*{0.228truein}


\vspace*{0.23truein}
\abstracts{It has been argued by several authors that
the quantum mechanical spectrum of black hole horizon area must be 
discrete. This has been confirmed in different formalisms, using
different approaches. Here we concentrate on two approaches, 
the one involving quantization on a reduced phase space of 
collective coordinates of a Black Hole and the algebraic approach 
of Bekenstein.  We show that for non-rotating, neutral
black holes in any spacetime dimension, the approaches 
are equivalent. We introduce a primary set of operators
sufficient for expressing the dynamical variables of both,
thus mapping the observables in the
two formalisms onto each other. The mapping predicts
a Planck size remnant for the black hole.
}{}{}

\vspace*{10pt}


\baselineskip=13pt	        
\normalsize              	
\section{What to Quantize?}	
\vspace*{-0.5pt}
\noindent
The central question of obtaining the theory of Quantum Gravity 
seems to be : what to ``quantize''. A perturbative approach has been
successfully attempted especially in the elegant work of B. DeWitt 
and  is adequate if one learns to contend
with the limitations of a nonrenormalizable theory. However the
nonperturbative aspects of the theory would remain inaccessible.
The formulation in terms of New Canonical Variables of Ashtekar
may be taken to be the minimal consistent nonperturbative approach 
to Gravity. It is a formulation amenable to solution on loop spaces,
originally pioneered by Mandelstam for QCD. Obtaining phenomenologically
interesting solutions however remains an unsolved problem.

This has spurred a number of other approaches wherein one assumes
certain ground states suggested by classical General Relativity
as possible vacuua. By focusing on a few collective coordinates
one attempts a quantization of these. Several approaches to 
Quantum Cosmology may be viewed in this light and seem to enjoy
success, again when interpreted with caution. In the following we shall
discuss such a procedure, formulated in \cite{bk,bdk}, 
applicable to any of the several Black Hole 
solutions assumed to be a given  ground state of the theory. This will
be referred to as reduced phase space quantization, or {\sl Approach I}.

A parallel approach to quantizing the collective coordinates of a
Black Hole is due to Bekenstein et al \cite{bm}. 
Unlike the canonical approach wherein
the canonical variables generate all the dynamical variables
relevant to a particular energy or length scale,  in the algebraic approach
one introduces each of  the operators by hand, guided by 
phenomenological observables. Quantization then amounts to ascertaining
all the commutation relations between the complete set of dynamical
variables\cite{dirac}. It is necessary in this approach that 
all the possible dynamical variables that are relevant at a particular
energy scale are consistently identified.  Cautious truncation
is necessitated within the complete list of dynamical variables of
a supposed complete theory. Finally,
in lieu of knowledge of the dynamics governing the system, one
relies on symmetries to propose a set of spectrum generating operators 
connecting the eigenstates of the observables. We shall refer to this
as the algebraic approach or {\sl Approach II}

Black Holes seem to present to us the happy situation where one 
may be reasonably confident that all the collective coordinates
of the system are known. Equivalently, being highly symmetric solutions
facilitate the task of the algebraic approach.
This has to do with the well known
``no hair'' property of the Black Holes. The only spoiler to this
seems to be the possibility that as one approaches the quantum
domain there may be phenomena occuring near the horizon which,
although inaccessible to the asymptotic observer, require additional
dynamical variables. We shall see that our ignorance of this can
however be encoded in appropriate operators ${\ha g}_{s_n}$.

A classical Black Hole is characterised by a short list of observables,
viz., electric and magnetic charge, angular momentum and the mass or 
equivalently the surface area of the horizon.
It has been argued by various authors, using widely
different approaches, that the spectra of above observables are
discrete
\cite{bk,bdk,bm,bek,kogan,Berezin,kastrup,louko,kastrup_strobl,VW,MRLP}.
In particular, the horizon area of a black hole has been shown to have
a uniformly spaced spectrum. 
Although the spectrum found in \cite{others} is not strictly uniformly spaced,
in the context of black hole entropy, as well as in a different
regularisation scheme \cite{poly}, the dominant contribution is equally
spaced.
In this talk we limit the discussion to show  the equivalence of
{\sl Approach I} and {\sl Approach II} for the case
of Schwarzschild black hole.
We propose a pair of primary operators $P$ and $P^\dag$ together
with a set of operators ${\ha g}_{s_n}$ (see eqn. (\ref{defn1})) which can
generate quantum algebras of both approaches consistently, thus 
implicitly mapping one model onto the other. While {\sl Approach II}
leaves open the value of the spectrum spacing, {\sl Approach I} 
predicts the unit of spacing, as also a zero point value for the same. 
We argue that such a zero point remnant is to be expected in  
{\sl Approach II} as well, and can be consistently included.

\section{The Reduced Phase Space}	
\noindent
It follows from the analysis
of \cite{dk,mk} that the dynamics of
static spherically symmetric configurations in
{\it any} classical theory of gravity in $d$-spacetime
dimensions is governed by an effective action of the form
\be
I = \int dt \le( P_M {\dot M} - H(M) \ri)
\ee
where $M$ is the mass and $P_M$ its conjugate momentum.
The above action can be rigorously derived by assuming
a generic gravity action (e.g. Einstein action, 
low energy string effective action or with or without 
cosmological constant), imposing spherical symmetry on it 
and doing a careful constraint analysis. However, intuitively 
it can be understood as follows: Birkhoff's theorem for uncharged 
spherically symmetric solutions of gravity states that the mass $M$
is the only time-independent and coordinate invariant parameter 
of the solution. That is ${\dot M} = 0$. The above effective action 
(with $H$ being independent of $P_M$)
is necessary and sufficient to guarantee this time independence.
The boundary conditions imposed are those of \cite{lw,shelemy}.
$P_M$ has
the interpretation of difference between the Schwarzschild times
between left and right infinities \cite{kuchar,thiemann,gkl}.

Now to restrict ourselves to black holes (and simultaneously 
exclude all other spherically symmetric configurations). Also,
as mentioned earlier, since the conjugate momentum $P_M$ 
playes the role of `time', 
motivated by Euclidean quantum gravity \cite{euclidean}, we assume 
that it is periodic with period which is
inverse the Hawking temperature ($T_H(M)$). That is,
\be
P_M \sim P_M + \f{\hbar}{T_H(M)} ~~.
\la{period}
\ee
This ensures that there is no conical singularity in the two 
dimensional euclidean
section near the black hole horizon. However, note that  
the above identification implies that the  physical phase space
is a wedge cut out from the full $(M,P_M)$ plane, bounded by the 
$M$ axis and the line $P_M=\hbar/T_H(M)$ \cite{dk,mk}.
Thus, we make the following canonical transformation
$(M,P_M) \rightarrow (X,\P_X)$, which
on the one hand `opens up' the phase space, and on the other hand,
naturally incorporates the periodicity (\ref{period}) \cite{bk,bdk}:
\ba
X &=& \sqrt{\f{A}{4\p G_d}} \cos \le( 2\p P_M T_H  /\hbar  \ri)
\la{ct1} \\
\P_X &=& \sqrt{\f{A}{4\p G_d}} \sin \le( 2\p P_M T_H /\hbar  \ri)
\la{ct2}
\ea
where $A$ is the black hole horizon area and $G_d$ the $d$-dimensional
Newton's constant. Note that both $A$ and $T_H$ are functions of $M$.
It can be shown that the validity of the first law of
black hole thermodynamics ensures that the above set of transformations
is indeed canonical \cite{bk,bdk}. Also
note that fixing the periodicity of $P_M$ to be $\hbar/T_H(M)$ uniquely 
fixes the prefactors in the right hand sides of (\ref{ct1}) and (\ref{ct2}).
Squaring and adding (\ref{ct1}) and (\ref{ct2}), we get:
\be
A = 4 \p G_d \le( X^2 + \P_X^2 \ri)~~.
\la{ham1}
\ee
The r.h.s. is nothing but the Hamiltonian of a simple
harmonic oscillator defined on the $(X,\P_X)$ phase space with mass $\m$
and angular frequency $\o$ given by $\m=1/\o=1/8\p G_d$.
Upon quantization, the `position' and `momentum' variables are
replaced by the operators:
\be
X \rightarrow {\ha X}~~~,~~~
\P_X \rightarrow {\ha \P_X} = - i \hbar \f{\pa}{\pa X}~~,
\la{op1}
\ee
and the spectrum of the black hole area operator follows immediately.
With $\ell_{pl}$ denoting the $d$-dimensional Planck length, and 
${\bar a} = 8\p \ell_{Pl}^{d-2}$, 
\be
A_n = {\bar a}(n\, +\, {1\over2})\ \equiv\  n {\bar a} + a_{Pl}~~~~~n=0, 1, 2, 
\cdots
\la{spec1}
\ee
Thus ${\bar a}$ signifies the basic quantum of area, and
$a_{Pl} = {\bar a}/2$ is its `zero-point' value.
Hawking radiation takes place when the black hole jumps
from a higher to a lower area level, the difference in quanta being
radiated away. The above spectrum shows that the black hole does not
evaporate completely, but a Planck size remnant is left over at the
end of the evaporation process.
It may be noted that the periodic orbits in the phase space under
consideration admit of an adiabatic invariant. As
mentioned earlier, in the present example,
the latter is in fact the horizon area of the black hole, just as it 
had been conjectured previously \cite{bm}. This follows from the integral
$$\mbox{Adiabatic Invariant} = \oint \P_X dX  = \f{A}{4G} ~~.$$

\section{Area as an Adiabatic Invariant}	
\noindent
Now let us consider {\sl Approach II}.  It has been argued that 
for a non-extremal black hole the area is an adiabatic invariant, 
and the spectrum emerges
from a proposed algebra of black hole observables\cite{bm}.
In the present work we take the case of neutral black hole in
zero angular momentum state. With slight modification of the notation
of \cite{bm} it is assumed that
there exists an operator ${\hat {\cal R}}_{ns_n}$ which
creates a single black hole state from vacuum $\vert 0 \rangle$
with area $a_n$ in an
internal quantum state $s_n$:
\ba
{\hat {\cal R}}_{ns_n}\vert ~0 \rangle &=& \vert n,s_n\rangle, \la{oa1}\\
{\hat A} ~\vert n, s_n\rangle &=& a_n ~\vert n, s_n\rangle~
\ea
We make the caveat that $s_n \in \{0,1, 
\ldots m_n-1 \}$ as in \cite {gour} such that the  
degeneracy of states with same area eigenvalue $a_n$, 
obeys $ \ln m_n \propto a_n$.  

Bekenstein introduced a minimal set of 
linear operators satisfying the following requirements:
(i) The commutator bracket 
between the operators must result in a linear 
combination of the operators in the set.
In other words, the algebra of black hole
operators must be linear and closed.
(ii) The area operator must commute with generators of
gauge transformations and rotations. This imposes
the physical requirement of invariance of the horizon 
area under these transformations.

The set of linear operators for the neutral black holes
will be area operator $\hat A$, black hole creation
operator ${\hat {\cal R}_{ns_n}}$ and its 
adjoint operator ${\hat {\cal R}}^\da_{ns_n}$ 
and identity operator $\hat I$. 
Bekenstein assumes that the vacuum state $\vert 0 \rangle$
has zero area in the construction of the algebra.
We shall denote Bekenstein's area operator as 
$\hat {A^{\prime}}$ with eigenvalues $a_n^{\prime}$ 
such that the vacuum area is $a_0^{\prime}=0$.
We will shortly see the relation between
the operators $\hat A$ and $\hat {A^{\prime}}$ 
and their respective eigenvalues $a_n$ and $a_n^{\prime}$.

With these requirements, Bekenstein's algebra for neutral
black hole will be \cite{bm}:  
\ba
{[}{\hat {A^{\prime}}}, {\hat {\cal R}}_{n s_n}] &=&a_n^{\prime}
{\ha {\cal R}}_{n s_n}~,
\la{al1}\\ 
{[} {\ha \car}_{ns_n} ,  {\ha \car}_{ms_m} ] &=& \epsilon_{nm}^k
{\ha \car}_{ks_k} ~~~(\epsilon_{nm}^k \neq 0~~{\rm iff}~ a_n^{\prime}+
a_m^{\prime} = a_k^{\prime})~,
\la {al2}\\
{[} {\hat {A^{\prime}}}, {[} {\ha {\cal R}}_{ms_m}^\da, {\ha {\cal R}}_{ns_n} ] ] &=&
(a_n^{\prime} - a_m^{\prime})
{[} {\hat {\cal R}}_{ms_m}^\da, {\hat {\cal R}}_{ns_n} ]~~{\rm if} ~
a_n^{\prime} > a_m^{\prime} ~, \nonumber \\
~&=&~0 ~~~~~~~{\rm otherwise} \la {al3}
\ea
Eqn. (\ref{al2}) implies that the black hole state created by
a commutator of two black hole creation operators (${\hat {\cal R}}_{ns_n}$,
${\hat {\cal R}}_{ms_m}$) will be another single
black hole state $\vert k, s_k\rangle$ provided its area 
satisfies $a_k^{\prime} = a_m^{\prime} + a_n^{\prime}$. 
Though the relation 
$$
[ {\hat {A^{\prime}}}, {\hat {\cal R}}^\da_{n s_n}] =-a_n^{\prime}
{\hat {\cal R}}^{\da}_{n s_n}
$$ 
is used to obtain eqn. (\ref{al3}), 
the positive definite nature of area operator $\hat {A^{\prime}}$
requires the inequality condition $a_n^{\prime} > a_m ^{\prime}$. 
Clearly, the spectrum of the above 
algebra $\{a_n^{\prime}\}$ involves both addition and subtraction of
area levels  which is possible if and only if the
area levels are equally spaced, i.e., 
\be
a_n^{\prime} = n \bar b,~~~~~n=0,1,2, \ldots~~
\ee
where $\bar b$ is some positive constant with dimensions of
area.

It is obvious that the neutral black hole algebra 
(\ref{al1} - \ref{al3}) is unchanged under the shift of the area operator:
\be
\hat {A^{\prime}} \rightarrow \hat {A^{\prime}} + {\bar c} 
\hat I~\equiv \hat A~,
\ee
where ${\bar c}$ is an arbitrary constant.
This relation between $\hat A$ and $\hat {A^{\prime}}$ 
implies their respective eigenvalues to satisfy
\be
a_n = a_n^{\prime} + {\bar c}~.
\ee 
Equivalently, the vacuum state  will have non-zero area $a_0=c$.
The situation is similar to the problems of single particle Quantum 
Mechanics where nontrivial zero-point energy always exists except for a 
free particle.  For the case of the
Hydrogen atom this is due to quantizing 
only the relative coordinates but not the coordinates of the centre 
of mass. In the case of the black  hole, the same is to be expected 
because we are not quantizing the  trivial collective  coordinates  
corresponding to its location.
The ${\bar c}$ must therefore be nonzero, presumably equal to $\bar b$ 
upto a dimensionless constant of order unity. If we identify $\bar b$
with the unit $\bar a$ obtained systematically in {\sl Approach I},
it is reasonable to also
identify $\bar c$ with $a_{Pl}={\bar a}/2= 4 \pi \ell_{p\ell}^{d-2}$. 

\section{Conciliation}	
\noindent
Our next step is to find a realisation of the operators in
{\it Approach II} in terms of the fundamental degrees of freedom
($M, \Pi_M$) in {\it Approach I}. 
We propose a representation of the algebra
(\ref {al1}-\ref{al3}) with the following form for the 
black hole creation operator ${\ha \car}_{ns_n}$ and 
area operator $\ha A$: 
\be
{\hat \car}_{ns_n} = (P^\da)^n~{\ha g}_{s_n}~;~~
\ha A = (\ha P^ \da \ha P + 1/2){\bar a}~,
\la{defn1}
\ee
where $\ha P^\da ~(\ha P)$ raises (lowers) the area level
$n$ to $n+1$ (respectively, $n-1$). The operators ${\ha g}_{s_n}$ 
 are similar to the secret operators in \cite {gour}. 
We postulate that these two sets of operators
satisfy the following commutation relations:
\ba
[\ha P, \ha P^\da] &=&1~, \la {pr1}\\
{[}\ha P, \ha g_{s_m}]&=&[\ha P^\da, \ha g_{s_m}]~
=~0, \la {pr2}\\
{[}{\hat g}_{s_m},{\hat g}_{s_n}] &=& \epsilon_{mn}^k {\hat g}_{s_k} 
~~~~{\rm where}~ \epsilon_{mn}^k \neq 0 ~{\rm iff}
~ k=m+n . \la {pr3}   
\ea
Equation (\ref{pr3}) ensures eqn. (\ref{al2}); however
it should be remembered that the operators $\ha g_{s_n}$
have a meaning only within the product form 
$(\ha P^\da)^n\ha g_{s_n}$.
Comparison with the reduced phase space approach 
(\ref{ct1}-\ref{spec1}) 
immediately gives us the form of $\ha P^\da$ as
\be
{\hat P}^\da = \f{1}{\sqrt{2\hbar} } \le[ {\hat X} - i {\hat \P_X}  \ri]~~.
\la{map1}
\ee
Note that the area operator (\ref{defn1}) becomes identical to 
that in {\sl Approach I}, namely Eq.(\ref{ham1}). 
The identification (\ref{map1}) shows that the black hole creation
operator ${\ha \car}_{ns_n}$ can be expressed in terms of fundamental
gravitational degrees of freedom $(M,P_M)$ via (\ref{ct1}),(\ref{ct2})
and (\ref{defn1}) in the following way:
\be
{\ha \car}_{ns_n}
= ({\ha P}^\da)^n~ {\hat g}_{s_n}~~;~~
{\ha P}^\da = \sqrt{\f{{\hat A}(M)}{8\p G_d \hbar}}~
\exp{\le( - i2\p {\ha P_M} {\hat T_H} (M)  \ri)}~~.
\ee
We see that the secret operator $\ha g_{s_n}$ in algebraic approach 
does not have a representation in terms of the fundamental 
gravitational degrees of freedom. This is consistent with the
no hair theorem where asymptotic observer cannot detect the
internal quantum state of the black hole. 

\section{Conclusion}	
\noindent
We have shown that approaches I and II are
equivalent in the zero angular momentum sector from the
asymptotic observer viewpoint, and hence
give rise to qualitatively similar spectra for the black hole area.
In {\sl Approach II}, in ref. \cite{bm}
the remnant (or zero-point) area was chosen to be zero. Relying on
single particle Quantum Mechanics experience we advocate
taking this to be non-zero; the presence
of the same in no way alters any of the commutators
(\ref{al1}) - (\ref{al3}). However note that the precise value of the
remnant remains undetermined in this approach. In the reduced phase space
approach on the other hand, the remnant is explicitly determined
to be a multiple of the Planck area in the relevant dimension. Since the
latter is the only natural length scale in quantum gravity, this seems
satisfactory. But a fundamental conclusion it suggests is that the
lowest energy state of the neutral black hole system is unique,
like the Hydrogen atom ground state.

Also note that the discrete spectrum (\ref{spec1})
means that Hawking radiation would consist of discrete spectrum
lines, enveloped by the semi-classical Planckian distribution.
As argued in \cite{bek,bk,bdk}, for Schwarzschild black holes of
mass $M$, the gap is order $1/M$, which is comparable to the frequency
at which the peak of the Planckian distribution takes place. Hence the
spectrum would be far from being a continuum, and can potentially be 
tested if and
when Hawking radiation becomes experimentally measurable.
Note that this is quite distinct from the predictions of loop 
quantum gravity, where it was shown that the resulting Hawking spectrum is 
practically continuous \cite{bcr}. 
It would also be interesting to explore the implications of the
Planck size remnant to the problem of information loss, since
the presence of the former can considerably influence Hawking Radiation
near the end stage of the black hole.

A further test of the correspondence elucidated in this article would
be to apply it to non-spherically symmetric as well as charged black
holes. Since both the approaches have dealt with electric charge,
analyzing the area and charged spectrum of a charged black hole
should be straightforward. However, it is to be borne in mind that
at least for semi-classical configurations (those with large
quantum numbers), the extremality bound has to be obeyed,
at least approximately. Incorporating angular momentum might be
somewhat tricky as the reduced phase space approach has not been
explored beyond the realm of spherical symmetry. We hope to report
on these and other related issues in the near future.

\nonumsection{Acknowledgments}
\noindent
SD would like to thank A. Barvinsky and G. Kunstatter for numerous
fruitful discussions and correspondences, and for collaboration,  
out which the reduced phase space approach emerged.  
The work of SD was supported in part by
the Natural Sciences and Engineering Research Council of Canada.
PR would like to thank the participants of  Gravity workshop, Ooty 
where some of the issues got clarified. UAY wishes to thank the 
organisers of the IGQR-2001 conference for the hospitality and the 
participants for stimulating discussions. Finally, we record our 
thanks to the referees for their scrutiny which resulted in improvement 
over the original version.

\nonumsection{References}

\end{document}